\documentclass[10pt,twocolumn,a4paper,superscriptaddress,prl,amsmath,amssymb,aps]{revtex4}
\usepackage{graphicx,graphicx,epsfig}

\usepackage{graphicx}
\usepackage{bm}

\begin{document}
\title{Geometric Control Over the Motion of Magnetic Domain Walls.}

\author{N.A. Sinitsyn}
\affiliation{CNLS/CCS-3, Los Alamos National Laboratory,
  Los Alamos, NM 87545 USA}
\author{V. V. Dobrovitski}
\affiliation{Ames Laboratory, Department of Physics and Astronomy, Iowa State University, Ames, Iowa 50011}
\author{S. Urazhdin}
\affiliation{Department of Physics,  West Virginia University, Morgantown, WV 26506}
\author{Avadh Saxena}
\affiliation{Theoretical Division, Los Alamos National Laboratory,
  Los Alamos, NM 87545 USA}

\begin{abstract}
We propose a method, which
enables precise control of  magnetic patterns, relying only on the fundamental properties of the wire and the choice of the path in the controlled parameter space
but not on the rate of motion along this path. Possible experimental realizations of this mechanism are discussed. 
In particular, we show that the domain walls in magnetic nanowires can be translated by rotation of the magnetic easy axis, or by applying pulses of
magnetic field directed transverse to the magnetic easy axis.
\end{abstract}

\date{\today} 

\maketitle

\begin{figure}[t]
\centerline{\includegraphics[width=8 cm]{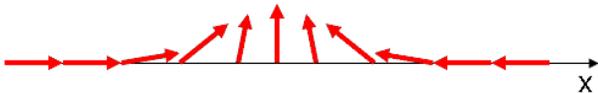}}
\caption{\label{wire1}  The domain wall between two domains with opposing magnetization directions parallel to the easy axis along the wire.
}
\end{figure}

Magnetic patterns such as domain walls, bubbles or skyrmions, can serve as information carriers in memory and logic devices~\cite{wallmemory}. To efficiently implement such devices,
 precisely controlled motion of the magnetic patterns must be achieved. Several traditional techniques have been
 employed, including magnetic fields~\cite{schryer}, field gradients~\cite{slonczewski-book} and electric currents~\cite{dwparkin}.
 These techniques create a gradient of the energy,
 which overcomes the pinning of the magnetic patterns on defects.  For example, displacement of a domain wall (DW) in a magnetic nanowire with the easy magnetic direction along the wire
(Fig.~\ref{wire1}) can be induced by applying a magnetic field pulse along the wire axis. The resulting displacement 
 of the DW depends sensitively on the amplitude and the duration of the pulse.

In this work, we demonstrate that the position of DW can be controlled {\it without} a gradient of the magnetic energy or voltage. Our approach is conceptually similar to applications of
 the Berry's geometric phases in topological quantum computing~\cite{artur-review}, and to the geometric mechanical control utilized, for example, for precise positioning of satellites
 \cite{marsden-book}. The general idea of geometric manipulations can be formulated as follows \cite{landsberg-92,wilczek-88}. Consider a system
exhibiting a continuous symmetry of steady states, such as invariance with respect to translation
 along some coordinate $x$. The system can be controlled by parameters $\vec{\mu}=\{\mu_1,\mu_2\ldots \mu_n \}$, variations of which affect the state of this
 system but do not break the symmetry. When such parameters are varied along a closed contour, the final state is generally different from the
 initial one by a shift $\delta x$ along the symmetry axis. This shift is purely geometrical, namely, it depends only on the choice of the contour in the
 controlled parameter space but in the adiabatic approximation does not depend on the rate of motion along  the contour. The geometric shift is given by
\begin{equation}
 \delta x= \oint \vec{A} \cdot d\vec{\mu},
\label{dx}
\end{equation}
where $\vec{A}$ is a vector function that depends on the parameter values.
 A well-known example of the geometric control in classical mechanics is a falling cat, who is capable to rotate its body into
 an upright position by periodic motion of its body parts.

Geometric control is robust against many errors in preparation of the controlling field pulses, which explains its success in applications to quantum computing and robotics. It is thus worth to explore 
the possibility of similar manipulations with patterns in ferromagnetic materials. 
We show below that the Landau-Lifshitz-Gilbert (LLG) equations describing the evolution
 of classical magnetic systems and the well known equations for slowly moving domain walls under external perturbations \cite{slonczewski-book}, lead to a simple realization of the geometric
 control over magnetic patterns such as
 180$^o$ domain walls in ferromagnetic nanowires. 

The geometric control can be implemented, for example, 
 by periodically varying the transverse magnetic field or the weak magnetic anisotropy perpendicular to the wire.
Unlike applied electric currents or magnetic field directed along the wire, periodic variation of these parameters results in the shift of the DW by a distance determined
only by the fundamental properties of the magnetic system, namely, the shift is robust with respect to fluctuations of the driving fields. However, pinning on impurities and the 
intrinsic Peierls-Nabarro barrier ~\cite{nabarro-dw} destroy translational symmetry,
 and present significant problem to demonstrate the effect. We estimate the strength of
a magnetic field pulse that can overcome it.  To demonstrate the effect, 
we separately consider two different directions of the magnetic easy axis with respect to the axis of the wire.

{\em DW in a wire with axial easy direction.} Consider a one dimensional magnetic wire with strong easy axis anisotropy along its axial direction $x$, and a weak easy axis anisotropy
 in the transverse direction forming an angle $\phi_0$ with the $y$-axis. The static magnetic energy is
\begin{equation}
\begin{array}{l}
E(x)=\int dx\{ J\left[(d\theta/dx)^2 +\sin^2 \theta (d\phi/dx)^2)\right]\\
\\
+K_0 \sin^2 \theta +K\sin^2\theta \sin^2(\phi-\phi_0) \},
\end{array}
\label{en}
\end{equation}
where $J$, $K_0$ and $K$ denote the exchange constant, the longitudinal anisotropy
and the weak transverse anisotropy strengths, respectively, $\theta=\theta(x)$ is the angle between the local magnetization vector and the axial direction $x$, and
$\phi$ is the angle between the projection of the magnetization on the $yz$-plane and the $y$-axis. We do not include the magnetic field or the long range dipolar interactions.
The dynamics of $\theta(x,t)$ and $\phi(x,t)$ are given by the LLG equations, which can be written in the Lagrange form
\begin{equation}
\frac{\delta L}{\delta X}-\frac{d}{dt} \left( \frac{\delta L}{\delta \dot{X}} \right) +\frac{\delta F}{\delta \dot{X}}=0,
\label{LL1}
\end{equation}
where $X$ is either $\theta(x)$ or $\phi(x)$, and
\begin{equation}
\begin{array}{l}
L=\int dx \{ E(x) +(M_s/\gamma) \dot{\phi}\cos \theta \},\\
\\
F=(\alpha M_s/2\gamma) \int dx \{ \dot{\theta}^2 +\dot{\phi}^2 \sin^2\theta \}.
\end{array}
\label{lag}
\end{equation}
Here, $F$ is the dissipation functional, $\gamma$ is the gyromagnetic ratio, $\alpha$ is the Gilbert damping parameter, and $M_s$ is the magnetization.
The DW  (Fig.\ref{wire1}) is the soliton solution of the stationary form of Eq.~(\ref{LL1}),
\begin{equation}
\theta(x)=2\tan^{-1}e^{(x-q)/\Delta}, 
\label{dw1}
\end{equation}
where $\Delta=\sqrt{J/K_0}$ is the width of the DW, and $q$ is its average coordinate.

We are interested in the DW dynamics induced by rotation of the weak anisotropy axis, which can be described by variation of $\phi_0$ in Eq.~(\ref{en}).
The simplest implementation of this rotation is to physically rotate the wire around its axis. We will discuss more practical implementations later. We assume that the rotation rate 
is much slower than the characteristic time scale $\sim 10^{-9}$~sec of the magnetic relaxation, allowing a transition from the original variables $\theta(x,t)$ and $\phi(x,t)$ to variables
 $q(t)$ and $\phi(t)$ that describe the position of the DW and its angle with the $y$-axis~\cite{schryer}. To derive the effective DW Lagrangian and the dissipation functional in these new
 variables, we substitute the DW solution (\ref{dw1}) into  (\ref{lag}), and integrate over $x$ while assuming that $q$ is time-dependent, yielding
\begin{equation}
\begin{array}{l}
L_{dw}=(2M_s/\gamma)\dot{\phi}q+2K\Delta \sin^2(\phi-\phi_0(t)),\\
\\
F_{dw}=\frac{\alpha M_s}{\gamma} \left( \frac{\dot{q}^2}{\Delta}+\dot{\phi}^2\Delta \right).
\end{array}
\label{lageff}
\end{equation}
The equations describing the DW dynamics are found by inserting Eq.~(\ref{lageff}) into Eq.~(\ref{LL1})
\begin{equation}
2K\Delta \sin\left( 2[\phi-\phi_0(t)] \right)-\frac{2M_s}{\gamma}\dot{q}+\frac{2\Delta M_s \alpha}{\gamma}\dot{\phi}=0,
\label{eqphi}
\end{equation} 
\begin{equation}
\dot{q}=-\Delta\dot{\phi}/\alpha.
\label{solq}
\end{equation}
In the adiabatic limit, the last two terms in Eq. (\ref{eqphi}) are negligible, yielding
$\phi \approx \phi_0(t)$. The coordinate of the DW is then determined from Eq. (\ref{solq})
\begin{equation}
q=q_0-\oint \frac{\Delta}{\alpha} d\phi_0=q_0-2\pi n\Delta/\alpha,
\label{qq}
\end{equation}
where $n$ is the number of rotations of the wire around its axis, and $q_0$ is the initial position of the DW.
The DW motion thus involves simultaneous rotation together with the anisotropy axis, and shift along the wire. Eq.~(\ref{qq}) directly corresponds to the general geometric control expression (\ref{dx}), where $\vec{\mu}$ is represented by the cyclically
 varied Cartesian components of the weak anisotropy axis, and   $A_{\phi_0}=-\Delta/\alpha$ is the connection that relates the evolution in the control
parameter space with the evolution in the "fiber space" represented by the DW coordinate $q$. For a steady rotation of the wire with angular frequency
 $\omega$, we find $q(t)=q_0-\frac{\omega\Delta }{\alpha} t$, i.e. the DW moves with a constant velocity $v=-\omega \Delta/\alpha$. Equation (\ref{qq})
 shows that the effect is geometrical in the sense that the displacement of the
DW depends only on the change in the control parameter $\phi_0$, but does not depend
on the rate of change as long as the adiabatic approximation is valid.
 Note also that the velocity does not depend on the strength of  transverse anisotropy $K$. Such robustness with respect to the  properties of the
 wire and the rate of the parameter variation is a general feature of the geometric control making it promising for practical applications.

It can be shown that the rotation of the wire can overcome the DW pinning as long as $\omega>H_c\gamma$, where $H_c$ is the critical axial magnetic field required to move the DW.
In permalloy Py=Ni$_{80}$Fe$_{20}$, $H_c\sim 1.5$~G~\cite{permalloy1}, yielding $\nu=\omega/(2\pi)= 4$~MHz, which is  too large for practical implementations.
 However, the equations of the DW motion derived above remain valid for any other mechanism driving the rotation of the  DW around the wire axis.  We note that the perturbation does not need to be steady. For example, one can use a short pulse of the circulating transverse magnetic field
\begin{equation}
H_y=H_{0y}(t)\cos(\omega t),\,\,\,\, H_z=H_{0z}(t)\sin (\omega t),
\label{hcirc}
\end{equation}
with amplitude sufficient to rotate the magnetization in the DW, and frequency $\nu> 4$~MHz. For the DW size $\Delta \sim 20$~nm and $\alpha \sim 0.01$~\cite{permalloy2}, a pulse
 of duration $\tau \sim 1$~$\mu$s can move the DW by a measurable distance $2\pi \tau \nu \Delta/\alpha \sim 0.06$~mm. The main challenge for the proposed experiment will be to create an 
rf field sufficient to overcome the transverse anisotropy of the nanowire, which for Py is dominated by the shape anisotropy. Standard nanofabrication techniques can produce
 approximately square Py nanowires with a
cross-section $50\times 50$~nm, and relative variations of the transverse dimensions
 not exceeding $20\%$. To overcome the resulting shape-induced anisotropy in Py then requires experimentally attainable $H_{0z}\approx 100$~Oe~\cite{pulse} and $H_{0y}$ - pulse amplitude along
transverse anisotropy axis can be much smaller.

{\em  DW in a wire with transverse easy direction.}
\begin{figure}[t]
\centerline{\includegraphics[width=8 cm]{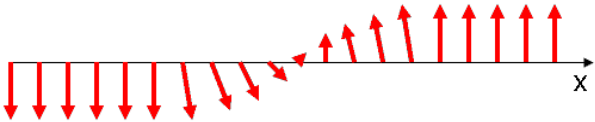}}
\caption{\label{wire2} The domain wall between two domains with opposing magnetization directions parallel to the easy axis transverse to the wire.
}
\end{figure}
Consider a 1D wire with a strong hard axis along the wire, and weak anisotropy in the transverse direction, described by the energy functional (\ref{en}) with $K_0 <0$, $K>0$, and $|K_0|> K$.
This anisotropy favors the magnetization direction in the transverse plane, as shown in Fig.~\ref{wire2}.
If it is sufficiently strong ($|K_0|\gg K$), then $\theta=\pi/2$ at any location along the wire.
The energy and  the dissipation functionals then can be simplified
\begin{equation}
\begin{array}{l}
E \approx \int dx \{ J(d\phi/dx)^2 + K \sin^2(\phi-\phi_0) \},\\
\\
F \approx \int dx \{\alpha M_s\dot{\phi}^2 / 2\gamma \}.
\end{array}
\label{DF-2}
\end{equation}
Without the driving fields ($\phi_0=const$),
the ground state is doubly degenerate at $\phi=\phi_0$ or $\phi=\phi_0+\pi$.
The DW solution connecting these two states is given by
\begin{equation}
\phi^{dw}(x;\phi_0,q)=\phi_0 + 2\tan^{-1}e^{(x-q)/\Delta}.
\label{dw2}
\end{equation}
The motion of this wall under circulating transverse magnetic field was predicted in \cite{coullet-91}. We will explore it
from the point of view of the geometric control Eq. (\ref{dx}).
The evolution equation of the angle $\phi(x)$ has a relaxation form
\begin{equation}
\frac{\alpha M_s}{\gamma} \partial_t\phi = 2J\partial_x^2 \phi -K\sin[2(\phi-\phi_0)]. 
\label{relax}
\end{equation}
 Assume that $\phi_0$ is slowly changed in
infinitesimal steps $\delta \phi_0$, allowing the system to relax after each such step. Assume that at some moment of time, the position of the DW is $q$, and the anisotropy direction is
 rotated from $\phi_0-\delta \phi_0$ to $\phi_0$.
Immediately after this rotation, the magnetic state ceases to be the stationary solution of Eq.~\ref{relax}.
 To find the new equilibrium state to which the DW will now relax, we linearize the evolution operator comprising the right-hand side of Eq.~(\ref{relax}) around
 the equilibrium solution (\ref{dw2}), taking the form $\hat{D}=2Jd^2/dx^2 -2K\cos[2(\phi^{dw}(x;\phi_0,q)-\phi_0)]$. The DW can be expanded in terms of the eigenstates $u_n(x)$ of this operator as
\cite{bouzidi-90}
\begin{equation}
\phi^{dw}(x;\phi_0-\delta \phi_0,q)=\phi^{dw}(x;\phi_0,q) + \sum \limits_n C_n u_n(x),
\label{dwe}
\end{equation}
where $C_n$ are time-dependent coefficients characterizing the evolution of the DW. Due to the relaxation, all the coefficients $C_n$ will eventually decay to zero, except for $C_0$ which corresponds to the zeroth mode of the operator $\hat{D}$, $u_0(x)=\partial_x\phi^{dw}(x;\phi_0,q)=\Delta/ \cosh \left[ (x-q)/\Delta \right]$ \cite{landsberg-92}.
Multiplying (\ref{dwe}) by $u_0(x)$, then integrating over $x$ and using the property $\int_{-\infty}^{\infty} dx u_0 (x) \phi^{dw}(x;\phi_0,q) =0$, we find
\begin{equation}
C_0 =\frac{-\delta \phi_0  \int_{-\infty}^{\infty} dx \{ u_0(x) \partial_{\phi_0} \phi^{dw}(x;\phi_0,q) \} }{\int_{-\infty}^{\infty} dx u_0^2(x)}=\frac{-\pi\Delta(\delta \phi_0)}{2}.
\label{c0}
\end{equation}
The new equilibrium state $\phi(x)=\phi^{dw}(x;\phi_0,q) +  C_0 u_0(x)$ is actually a new DW solution shifted by $\delta q =-C_0 $. This can be shown by expanding the shifted DW solution as $\phi^{dw}(x;\phi_0,q+\delta q) \approx \phi^{dw}(x;\phi_0,q)-\partial_x\phi^{dw}(x;\phi_0,q)\delta q$, and noting that $u_0(x)=\partial_x\phi^{dw}(x;\phi_0,q)$. For cyclic evolution of the control parameter, the geometric shift now reads
\begin{figure}[t]
\centerline{\includegraphics[width=7 cm]{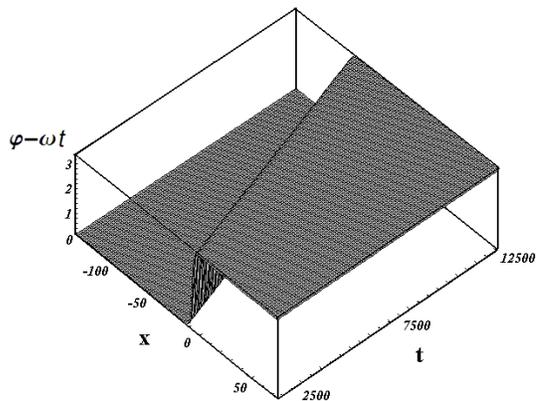}}
\caption{\label{DW1} Evolution of the profile of the DW $\phi(x,t)-\omega t$. The parameters used in the calculation are $\frac{\alpha M_s}{2\gamma}=1$,
$\phi_0=\omega t$, $\omega= -0.00035$, $K=1$, $J=1$, $H_y=0$. Time $t$ and coordinate $x$ are in arbitrary units.
}
\end{figure}
\begin{equation}
\delta q = \oint A_{\phi_0}d\phi_0=n\pi^2 \Delta, \,\,\,\,\,\,\, A_{\phi_0} = \pi \Delta/2. 
\label{shift-2}
\end{equation} 
In contrast to the case of wires with axial easy direction, the shift is independent of the Gilbert damping $\alpha$ as long as the perturbation is adiabatic. Fig.~\ref{DW1} shows the numerical solution of Eq.~(\ref{relax}) demonstrating
 the DW motion consistent with Eq.~(\ref{shift-2}).

The forces driving the DW due to the rotation are suppressed by a small parameter $\dot{\phi_0} \tau$, where $\tau\sim 10^{-8}-10^{-9}s$ is 
typical magnetization relaxation time. They are considerably smaller in this geometry than pinning forces. 
Hence, rotation alone is not sufficient. However, in combination with a strong periodic perturbation sufficient to overcome the pinning, the
geometric shift can become important. 
The strong perturbation can produce a zero effect on average,
 while the small geometric shifts accumulate into a significant net displacement.
 A well-known analogy of this configuration in classical mechanics is the Foucault pendulum.
 Its fast dynamic oscillations average to zero, while the geometric angle caused by the Earth rotation slowly accumulates.

An example of such a perturbation is provided by a fixed transverse magnetic field $H_y$,
 which is much weaker than the effective transverse anisotropy field, but sufficient to overcome the DW pinning.
 We consider the same rotating wire, whose energy  (\ref{DF-2}) is now modified according to
\begin{equation}
E'= E-H_yM_s\int  dx   \cos (\phi).
\label{DF-3}
\end{equation}
Solving Eq.~\ref{LL1} with a constant $\phi_0$, we obtain the field-induced DW velocity
\begin{equation}
\dot{q} \approx \gamma H_y \Delta \cos(\phi_0)/\alpha.
\label{dotq}
\end{equation}
In the rotating wire frame, the magnetic field becomes oscillating. According to Eq.~\ref{dotq}, it moves the DW in one direction during one half of the rotation cycle, and in the opposite direction during the other half.
These oscillations are superimposed on a weak but accumulating geometric effect due to the rotation of the anisotropy axis.
\begin{figure}[t]
\centerline{\includegraphics[width=7 cm]{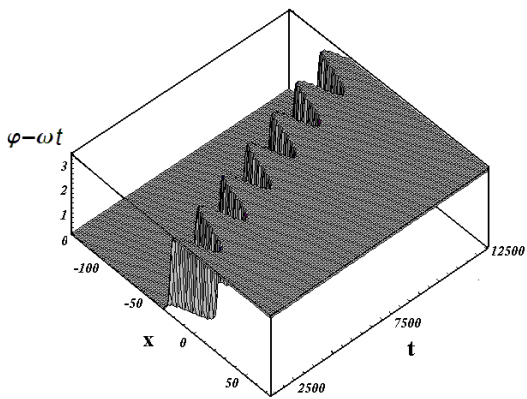}}
\caption{\label{DW2} Evolution of the DW profile for the same parameters as in Fig.~\ref{DW1} in an additional magnetic field $M_sH_y=0.012$.
} 
\end{figure}
Fig.~\ref{DW2} shows the results of simulations for the same system as in Fig.~\ref{DW1}, but in the presence of the transverse magnetic field. DW oscillation due to the magnetic field is dominant during each rotation period, but the accumulation of geometric shifts becomes apparent after several full rotations. This effect can possibly contribute to
 purification of magnetic samples from topological defects (not only domain walls) during sufficiently long rotation in an external magnetic field.


{\em In summary}, we demonstrated that the problem of a classical dissipative domain wall motion contains a useful geometric structure enabling new schemes for robust magnetic manipulations.
We discussed two examples
of the DW geometric shift in ferromagnetic wires with different directions of magnetic anisotropy.
We showed that adding a periodic perturbation can alleviate the potentially detrimental effect of pinning on impurities. The concept of geometric control is
not restricted to the two proposed models, and our work should stimulate efforts to find its practical implementations in other magnetic
materials.
Other magnetic patterns, such as spin-flop states, spiral waves, and skyrmions, may lend themselves to similar geometric manipulation schemes.
 Oscillating perturbations have been studied in many systems described by the Ginzburg-Landau type of equations, including domain walls motion in liquid crystals
\cite{slonczewski-book,frisch-94,coullet-91,rudiger-07,kawagishi-95,vierheilig-97}.
It may be possible to similarly interpret many of these results in geometric terms, as expressed by Eq. (\ref{dx}).
Finally, it would be
interesting to explore the geometric domain wall shifts in systems with other symmetries of the order parameter, e.g. in ferroelectrics and antiferromagnets.


{\em Acknowledgments.}
  This work was funded in part by DOE under Contract No.
  DE-AC52-06NA25396, and NSF Grant DMR-0747609.

\end{document}